\documentclass[prr,onecolumn,superscriptaddress]{revtex4}

\usepackage{amssymb,amsbsy}
\usepackage{graphicx}
\usepackage{rotating}
\usepackage{dcolumn}
\usepackage{amsmath}
\usepackage[abs]{overpic}

\newcommand{\abs}[1]{\vert #1 \vert}

\begin{document}
\title{An information theoretic network approach to socioeconomic correlations}

\author{Alec Kirkley}
\affiliation{Department of Physics, University of Michigan, Ann Arbor, Michigan 48109, USA}

\begin{abstract}

Due to its wide reaching implications for everything from identifying hotspots of income inequality to political redistricting, there is a rich body of literature across the sciences quantifying spatial patterns in socioeconomic data. In particular, the variability of indicators relevant to social and economic well-being between localized populations is of great interest, as it pertains to the spatial manifestations of inequality and segregation. However, heterogeneity in population density, sensitivity of statistical analyses to spatial aggregation, and the importance of pre-drawn political boundaries for policy intervention may decrease the efficacy and relevance of existing methods for analyzing spatial socioeconomic data. Additionally, these measures commonly lack either a framework for comparing results for qualitative and quantitative data on the same scale, or a mechanism for generalization to multi-region correlations. To mitigate these issues associated with traditional spatial measures, here we view local deviations in socioeconomic variables from a topological lens rather than a spatial one, and use a novel information theoretic network approach based on the Generalized Jensen Shannon Divergence to distinguish distributional quantities across adjacent regions. We apply our methodology in a series of experiments to study the network of neighboring census tracts in the continental US, quantifying the decay in two-point distributional correlations across the network, examining the county-level socioeconomic disparities induced from the aggregation of tracts, and constructing an algorithm for the division of a city into homogeneous clusters. These results provide a new framework for analyzing the variation of attributes across regional populations, and shed light on new, universal patterns in socioeconomic attributes.    
\end{abstract}
\maketitle

\section{Introduction}

Analysis of spatial data is crucial for understanding a wide variety of human systems, and with our increasing capacity to handle high resolution data sets, has found applications across the sciences \cite{bailey1995interactive}. From analyzing demographic polling behavior \cite{johnston1979political}, to epidemic vulnerability of populations \cite{elliot2000spatial}, to disparities in access to nutritious food \cite{walker2010disparities}, spatial data on social and economic attributes of populations is central to many problems in modern data-driven science. In particular, assessing the extent to which socioeconomic properties differ across regions of space is an important topic for understanding the spatial dynamics of production and consumption \cite{beckmann1985spatial}, the manifestation of segregation \cite{reardon2004measures}, and the spatial decomposition of inequality \cite{rey2004spatial}. There has been thus been extensive research to understand how socioeconomic indicators fluctuate across space, which has involved the development of many sophisticated mathematical techniques to quantify variation in spatial data. \\

A major challenge with these methods is determining the scale at which to probe spatial variations, noting that populations tend to disperse heterogeneously across space \cite{brown1995spatial}. Extreme spatial inhomogeneity in population density causes inconsistencies in interpretability of results based solely on distance, as the pace of economic activity is largely determined by population \cite{bettencourt2007growth}, and space is primarily relevant insofar as it relates to the number of ``intervening opportunities" it provides economic agents with  \cite{stouffer1960intervening}. As a result, various methods have been designed to account for heterogeneous population in the analysis of spatial data, some of which include density-equalizing maps \cite{gastner2004diffusion} and methods based on dasymetric mapping \cite{holt2004dasymetric}. As an additional complication, there is no apriori way to aggregate regions of space for statistical analysis (an issue that is more precisely quantified by the Modifiable Areal Unit Problem, or MAUP for short) \cite{gehlke1934certain}. Consequently, finding suitable scales for various problems in spatial analysis is an open problem that has received extensive interest due the sensitivity of results to the chosen scale \cite{turner1989effects}. To make matters worse, policy interventions take place at the level of artificial government-designated boundaries, and so analysis that ignores these boundaries may be irrelevant for certain implementation-driven studies \cite{flint2007political}. Here, we assess relationships between official boundaries (census tracts) in a network-based manner, circumventing the aforementioned spatial issues by considering topological distance rather than geographic distance, which also allows for the development of insights at the scale of regions designated for policy intervention.  \\

Another important avenue of research investigates what measures to use to quantify spatial disparities in population data. For nominal distributions, such as race or religious affiliation, there are a wide variety of measures of qualitative variation that take the form of segregation indices between disjoint groups \cite{duncan1955methodological}. Some of these indices can also be used with ordinal or interval data such as population counts over income brackets or education levels \cite{reardon2004measures}, but few of these measures have the flexibility to accommodate all types of data on the same scale or generalize to more than two regions. Measures based on information theory can also be effectively applied to distributional socioeconomic data \cite{walsh1979information,chodrow2017structure}, having the additional benefit of being founded in fundamental statistical principles, and allowing in some cases an extension to multiple distributions. We develop here a novel approach based on the Generalized Jensen-Shannon Divergence (GJSD) \cite{lin1991divergence} to compare distributional data, which has a number of advantages over other approaches, including flexibility for all distributional data types and an intuitive theoretical interpretation.   \\

We note other approaches proposed to analyze spatial data using networks or information theoretic principles, as there has been similar research in regional science, economic and political geography, urban planning, and spatial analysis. There has been extensive work on using spatial network methods in urban science \cite{barthelemy2011spatial,kirkley2018betweenness}, with a focus oriented mostly towards the structure of urban form and the dynamics of urban growth. Numerous methods based on spatial aggregation of neighboring regions within the context of multiscalar diversity indices have been developed to assess the spatial manifestation of diversity \cite{clark2015multiscalar,olteanu2019segregation}, but rarely do these accommodate distributions or multiple data types. Additionally, there is a body of research constructing spatial correlation and aggregation methods based on information theoretic measures \cite{batty1974spatial,vaz2018merging}. However, these analyses thus far have been limited primarily to racial segregation and ecological diversity, and focus on relationships between individual regional entities within larger clusters rather than multi-distribution correlations as in this study. Additionally, these measures may not be as easily interpreted in terms of a simple statistical process, as is the case with our method, and many of these measures are not adaptable to all data types, which limits their capability in comparative analysis. \\

In this study, we develop a novel approach for studying spatial variation in distributional socioeconomic data based on regional adjacency networks and information theory, and apply our methodology to the network of adjacent census tracts in the continental US through a few experiments. We first examine two-point correlations in our distributional distance measure with respect to path length across the adjacency network, finding a universal decay pattern with similar scaling exponents and finite size cutoffs across a variety of socioeconomic attributes. We also utilize this methodology to assess disparity with respect to various socioeconomic attributes across US counties by generalizing our measure to the comparison of more than two regions, finding high regional dependence and correlations in our measure for multiple variables. Finally, we discuss a new means for spatial aggregation of regions through community detection with our measures at multiple resolutions, clustering the census tract network for the city of Chicago into meaningful regions of homogeneous socioeconomic characteristics at different cluster size scales. Our methods provide a new means for analyzing spatial variation in all types of distributional data within a universal framework that circumvents limitations in traditional spatial measures. The results from our experiments point to new ways of thinking about how socioeconomic characteristics manifest across space, and can be applied to a wide range of problems across the social sciences.                 

\section{Methodology}

\subsection{Census Tract Data and Network Construction}

In order to study a wide variety of socioeconomic attributes at high spatial resolution, we utilize US Census data at the tract level from the American Community Survey in 2018 \cite{us2004american,acsurl}. The American Community Survey continuously samples US households to collect data on various socioeconomic and demographic characteristics of the population, and it is the largest survey at the household level that is conducted by the Census Bureau. We choose to analyze data at the level of census tracts because they encapsulate highly localized populations, represent officially designated regions relevant for policy intervention \cite{houstoun1976neighborhood,krieger2006century}, and have roughly equal populations (the $25$th and $75$th percentiles in terms of population are $2971$ and $5572$ for the set of tracts used in the analysis). We aggregate distributional data on educational attainment, house price, income, industry of occupation, and race in order to assess spatial variability across a range of different variables. The techniques we develop can be adapted for continuous distributional data, but here we use the available binned data for housing prices and incomes, leaving to future work the estimation of the full corresponding continuous distributions, as this is a difficult problem on its own \cite{mcdonald2008generalized,von2017better}. \\

In order to quantify variation in the discrete distribution of a variable $X$ across tracts, we encode its possible values as a vector $q_X$, which may be nominal, ordinal, or interval in nature depending on the variable $X$ being analyzed. For census tract $i$, we denote its distributional vector of values for the variable $X$ as $q^{(i)}_X$, with the particular value for an entry $x$ denoted $q^{(i)}_{X}(x)$. These tract distributional vectors are normalized, and satisfy $\sum_{x}q^{(i)}_{X}(x)=1$ for all tracts $i$ and variables $X$, making $q^{(i)}_{X}(x)$ a probability mass function over realizations $x$ of $X$. For example, if in census tract $5$ there are $300$ persons classified as \emph{Asian} out of $1000$ total persons, then $q^{(5)}_{race}(Asian)=0.30$. Details on the variables analyzed are given in Table I. \\

The nearest-neighbor network representation for census tracts is constructed utilizing the TIGER shapefile data \cite{united2019tiger}, and two tracts are neighbors in the network if they share a common length of border or a corner. Only tracts in the continental US were considered for this analysis in order to ensure a single connected component for two-point correlation analyses. After removing tracts with corrupted or incomplete data, the final network had $70,201$ nodes and $197,841$ edges (for an average degree of $5.6$). The overall goal in terms of practical relevance of the proposed methods is for local spatially targeted interventions (e.g. at the scale of counties, cities, or neighborhoods, with tracts as the fundamental subdivision), and so we are only presently interested in relatively short range correlations, hence the choice to construct the underlying network based on spatial adjacency. However, the method we present for comparing local distributions of socioeconomic variables can be applied to any pair of regions (whether or not they are adjacent), which is in fact what is done for our two-point correlation analysis, and so any network structure signifying a relationship between two regions could be used in this framework. For instance, one could construct the network based on population migration flows from region to region, which would no longer necessarily have geographically localized edges, but could be used to see whether or not people move homes between regions with similar or different socioeconomic properties.\\

As the analyses performed in this study are intentionally topological in nature, rather than geographic, we do not focus on spatial dimensions. However, for better contextualization of our results for those unfamiliar with the spatial extent of subdivisions within the US, we report summary statistics from our network dataset here. For the subset of tracts used in the analysis, the distribution of land areas is heavily right-skewed, with the tracts in the $10$th percentile, median, and $90$th percentile having areas of $1.0\;\text{km}^2$, $6.4\;\text{km}^2$, and $269.2\;\text{km}^2$ respectively. If we consider the set of tracts kept in the filtered dataset, and construct their (potentially incomplete) associated counties, the distribution of land areas is also right-skewed, with the counties in the $10$th percentile, median, and $90$th percentile having areas of $953.0\;\text{km}^2$, $1911.4\;\text{km}^2$, and $4863.0\;\text{km}^2$ respectively. The high level of heterogeneity we see in the land area statistics at both the tract and county level further illustrates the utility of an approach to socioeconomic correlations that is spatial scale-independent, as adminstratively equivalent regions clearly can have drastically different sizes.

\subsection{Generalized Jensen-Shannon Divergence}

Due to its desirable properties as a distributional distance measure, which we discuss in more detail, the Generalized Jensen-Shannon Divergence (GJSD) has gained popularity for applications across disciplines, from quantum physics \cite{briet2009properties}, to genomics \cite{itzkovitz2010overlapping}, and even to history \cite{klingenstein2014civilizing}. For our purposes, the GJSD will allow us to distinguish distributional data across census tracts in a meaningful way, which can be understood in terms of the following process. \\

Suppose we have two spatial regions, region $1$ and region $2$, and we want to determine how similar these regions are with respect to a socioeconomic variable $X$. We assume that their respective populations $n_1$ and $n_2$ are known, as well as the distributions $q^{(1)}_{X}(x)$ and $q^{(2)}_{X}(x)$ defined in the Introduction. One way to think about how the populations in regions $1$ and $2$ differ in their composition of the attribute $X$ is to consider the situation where there was no artificial line drawn between regions $1$ and $2$, and instead we had just decided to consider them one single  ``super-region". We can then ask the question: How different is the distribution of $X$ across the population in this super-region than in its individual sub-regions? Rather than naively comparing the distributions $q^{(1)}$ and $q^{(2)}$ directly, this perspective accounts for the population difference between the regions, and will also allow us to address in a natural way the increase in regional homogeneity we get by separating these regions. \\

From an information theoretic perspective, we can quantify the homogeneity of attribute $X$ within a population by its \emph{average information content} (or \emph{surprisal}), in other words how unpredictable it is. For instance, if a population has relatively equal fractions of people from each race, it is difficult to guess what any given person's race is, and the amount of ``information" we gain by finding out each person's race is relatively high on average. However, if nearly everyone is of a single race, it is very easy to guess an individual's race, and we are on average very ``unsurprised" upon each discovery of the race of a randomly selected individual in this population. For our thought experiment, we can determine the homogeneity gain we achieve by separating regions $1$ and $2$ by computing how much the average information content of attribute $X$ in the population is reduced after the split of the super-region. \\

The average information content of a random variable with probability distribution $q(x)$ is given by its entropy, $H[q(x)]$, where $H$ is the Shannon entropy functional
\begin{align}
H[q(x)] = -\sum_{x}q(x)\log q(x)  
\end{align}
and $\log q(x)$ is the information content of an observation of $x$ \cite{cover2012elements}. Thus, the average information content of attribute $X$ in the super-region population is given by
\begin{align}
H\left[Q^{(12)}_X(x)\right] = -\sum_{x}Q^{(12)}_X(x)\log \left(Q^{(12)}_X(x)\right),    
\end{align}
where
\begin{align}
Q^{(12)}_X(x) = \frac{n_1}{n_1+n_2}q^{(1)}_{X}(x)+\frac{n_2}{n_1+n_2}q^{(2)}_{X}(x)
\end{align}
is the empirical probability mass function of $X$ in the super-region. Now, if regions $1$ and $2$ are split, then we can associate to any individual in the super-region a label $i=1,2$ denoting the region they are from, which will necessarily reduce our uncertainty about their value of $X$ on average. Then, in a random experiment to survey the same population about $X$, we would know the region $i$ that each person we sample is from, and thus the information content associated with each observation we make is $\log q^{(i)}_{X}(x)$ rather than $\log Q^{(12)}_X(x)$. The average information content $H'\left[Q^{(12)}_X(x)\right]$ of $X$ after the regional split is then given by the weighted average
\begin{align}
H'\left[Q^{(12)}_X(x)\right] &= \frac{n_1}{n_1+n_2}H[q^{(1)}_X(x)]+\frac{n_2}{n_1+n_2}H[q^{(2)}_X(x)], 
\end{align}
and the reduction in average information content from splitting the regions is given by the difference
\begin{align}
J^{(12)}_{X} = H\left[Q^{(12)}_X(x)\right] - H'\left[Q^{(12)}_X(x)\right].   
\end{align}

Generalizing our argument to the merging of $m\geq 2$ regions, we have that the reduction in average information content by the separation of $m$ adjacent regions is given by
\begin{align}
\label{J_general_def}
J^{(1,...,m)}_X = H\left[Q^{(1,...,m)}_X(x)\right] - \sum_{k=1}^{m}\pi_kH[q^{(k)}_X(x)],    
\end{align}
where
\begin{align}
\pi_k = \frac{n_{k}}{\sum_{k'=1}^{m}n_{k'}}   
\end{align}
(with $n_k$ the population of region $k$) and
\begin{align}
Q^{(1,...,m)}_X(x) = \sum_{k=1}^{m}\pi_kq^{(k)}_X(x).    
\end{align}
We can recognize now that Eq. \ref{J_general_def} is equivalent to the \emph{Generalized Jensen-Shannon Divergence} (GJSD), which is sometimes referred to as just the \emph{Jensen-Shannon Divergence} for $m=2$ \cite{lin1991divergence}.\\

Intuitively, if the distributions $\{q^{(k)}\}$ are all very similar, knowing which region that a person is from does not reduce our uncertainty about their value of $X$ by much, and $J^{(1,..,m)}$ will be close to $0$. On the other hand, if the $\{q^{(k)}\}$ are relatively different, then we can reduce our uncertainty about a person's value of $X$ by a lot by knowing which region $k$ they are from, and $J^{(1,...,m)}$ will be higher.\\  

We know that Eq. \ref{J_general_def} is bounded below by $0$ due to the concavity of entropy, and this minimum is achieved when $q^{(k)}=q^{(k')}$ for all $k,k'$, as merging the regions does not change our uncertainty about a person's value of $X$ at all. On the other hand, the maximum value $J^{(1,..,m)}_{max}$ that Eq. \ref{J_general_def} can take is
\begin{align}
\label{max_J}
J^{(1,...,m)}_{max} = -\sum_{k=1}^{m}\pi_k\log \pi_k,      
\end{align}
which happens when the $\{q^{(k)}\}$ are entirely non-overlapping in their regions of non-zero probability. We can see that this is the upper bound by rewriting Eq. \ref{J_general_def} in a more illuminating manner as
\begin{align}
J^{(1,...,m)}_X = \sum_{x,k}\pi_kq^{(k)}(x)\log\left[\frac{q^{(k)}(x)}{\sum_{l}\pi_{l}q^{(l)}(x)}\right],
\end{align}
and noting that $\log\left[\frac{q^{(k)}(x)}{\sum_{l}\pi_{l}q^{(l)}(x)}\right]\leq \log\left[\frac{1}{\pi_k}\right]$, with the equality holding when $q^{(l)}(x) = 0$ for all $l\neq k$, which is equivalent to the $q$'s having disjoint nonzero support. Eq. \ref{max_J} is just the average uncertainty we have about which smaller region $k$ a randomly chosen person from the super-region will come from.\\

We normalize Eq. \ref{J_general_def} by the upper bound in Eq. \ref{max_J} to enforce values to lie in $[0,1]$, which allows us to compare tract similarities for regions with variable populations $n_k$. The final expression we use for distributional comparison across regions is then
\begin{align}
\label{L_eq}
L^{(1,...,m)}_{X} = \frac{J^{(1,...,m)}_{X}}{J^{(1,...,m)}_{max}}. 
\end{align}
This measure is easily adapted to any discrete variable $X$, which can be nominal, ordinal, or interval in nature, allowing for the application of Eq. \ref{L_eq} to a wide variety of problems. It can also be adapted to continuous distributions through approximations of the differential entropy. We note that for ordered data, Eq. \ref{L_eq} is only sensitive to how much the probability mass changes between distributions of interest, not to where it moves. In this sense, there are other appealing measures for comparing ordered data, such as variants of the earth-mover's distance \cite{rubner1998metric}. However, Eq. \ref{L_eq} has a major advantage over such previous measures in that it can be used to compare results across all types distributional data on the same scale, and can also accommodate the inclusion of more than two distributions for comparison. In the following section, we perform multiple experiments on the tract adjacency network using Eq. \ref{L_eq}, demonstrating new insights on spatial socioeconomic variability that can be gained through our methodology. 

\section{Results}

\subsection{Two-point Correlations in $L_X$}
Two-point correlation functions---a term used to refer generically to functions that measure some type of average correlation between points in a system as a function of the distance between them---are an invaluable tool for describing spatial data for systems as diverse as galaxy clusters \cite{davis1983survey}, turbulent fluids \cite{ganapathisubramani2005investigation}, and earthquakes 
\cite{kagan1980spatial}. In more recent work, the concept of the two-point correlation function has been extended to networks \cite{rybski2010quantifying,mayo2015long,fujiki2018general}, where it refers to computing correlations between the properties (in most cases, degree) of two nodes as a function of the shortest path distance between them. \\

Here, in order to assess the ``scale" at which socioeconomic properties vary across the US, we compute a two-point correlation function for $L_X$ (Eq. \ref{L_eq}) between census tracts as a function of the number of network hops between them. The effective distance we are concerned with is then consistent with policy-relevant boundaries and roughly accounts for the heterogeneous population density across space (as tracts have relatively similar populations as discussed earlier). In other words, the total population of neighbors at path distance $l$ or less from a focal tract is roughly the same for all tracts, as the degree distribution of the analyzed network is highly homogeneous as is characteristic of spatial networks in general.\\

In our case, the two-point correlation function $C_X(l)$ for socioeconomic attribute $X$ as a function of (unweighted) network geodesic distance $l$ is given by
\begin{align}
\label{twopt}
C_X(l) = \frac{1}{n(l)}\sum_{i<j}L_X^{(ij)}\delta_{d_{ij},l}
\end{align}
where $\delta$ is the Kronecker delta function, $n(l)$ is the number of node pairs separated by shortest path distance $l$, and $d_{ij}$ is the shortest path distance between tracts $i$ and $j$ in the adjacency network. $C_X(l)$ gives the average divergence $L_X^{(ij)}$ over all pairs of nodes $(i,j)$ that are separated by $l$ hops. \\

Calculating $C_X(l)$ exactly is difficult, as there are $\sim 2.5$ billion pairs of tracts $\{i,j\}$ in the network that contribute to the sum in Eq. \ref{twopt} for a given $X$. We therefore opt for a sampling procedure to compute $C_X(l)$ approximately, traversing nodes $j$ in the network up to a distance $l=20$ starting at $1,000$ uniformly sampled focal tracts $i$, then computing the sum in Eq. \ref{twopt} over sampled focal tracts $i$ and traversed nodes $j$. A network distance of $l=20$ corresponds to a spatial distance of $~200$ km, varying depending on the location of the central tract, and so captures spatial regions roughly of size $160,000~\text{km}^2$ (or about $2\%$ of the land area of the continental US). Using this distance cutoff thus restricts our analysis to relatively concentrated regions, which may be more relevant for spatially targeted policy interventions.   \\

In order to examine the scale over which correlations in each attribute decay, we analyze how quickly $C_X(l)$ approaches its asymptotic value $C_X(\infty)$ from its initial value $C_X(1)$ as we increase $l$. $C_X(\infty)$ is estimated as the average value of $L_X$ over 10,000 tract pairs selected uniformly at random (which should draw primarily from node pairs at distances much greater than $l=20$ based on the network structure). Taking inspiration from the form of two-point correlations in spin systems, we can then fit the resulting data to the truncated power-law form 
\begin{align}
\label{pl_form}
\tilde C_X(l) = l^{-\alpha}e^{-(l-1)/\beta},    
\end{align}
where
\begin{align}
\tilde C_X(l) = \frac{C_X(\infty)-C_X(l)}{C_X(\infty)-C_X(1)},   
\end{align}
and we've rescaled $C_X\to \tilde C_X$ to account for the intercepts at $l=1$ and $l=\infty$. \\

The scaling exponent $\alpha$ in Eq. \ref{pl_form} quantifies the rate of decay in correlation in the system as a function of distance (network hops), and the cutoff exponent $\beta$ determines the distance scale (in terms of hops) over which correlation persists. A higher (more positive) value of $\alpha$ indicates a slower decay in correlations as we move away from a given tract, and a higher value of $\beta$ indicates a longer distance over which tracts have correlated distributions with this reference tract. To extract the exponents $\alpha$ and $\beta$, the following OLS fit is performed 
\begin{align}
\label{fit}
\log \tilde C_X(l) = -\alpha \log l - \frac{(l-1)}{\beta}+\epsilon_l,
\end{align}
with $\epsilon_l$ a white noise process.  \\

We plot the results of the fit in Fig. 1A, where we show the coefficient of determination $r^2$, the scaling exponent $\alpha$, and the cutoff exponent $\beta$ for the fit in Eq. \ref{fit} for each attribute. We can see that the curves for all attributes (apart from ``industry", which due to autocorrelated residuals has been suspected to follow a different decay form that we will not explore here) collapse quite well onto each other. This collapse is not only an indication of a good fit, but can possibly lead us to consider a more fundamental, attribute-independent mechanism behind the variation of different attributes $X$ across regions, which we will discuss at the section's closing. \\

To investigate a potential consequence of the striking similarity in the decay of $\tilde C_X(l)$ across attributes $X$ studied in Fig. 1A, we examine the correlations between the losses $L_X^{(ij)}$ and $L_{X'}^{(ij)}$ across edges $(i,j)$ for all pairs of attributes $(X,X')$. Specifically, we analyze the monotonic dependence between losses using Spearman correlation, which relaxes the linearity assumption of Pearson correlation but also allows us to test for the significance of observed correlations \cite{gautheir2001detecting}. Specifically, we compute
\begin{align}
\label{rho}
\rho\left(\{L_X^{(ij)}:(i,j)\in E\},\{L_{X'}^{(ij)}:(i,j)\in E\}\right),
\end{align}
where $E$ is the set of edges in the adjacency network, $\rho$ is the Spearman correlation coefficient, and the arguments to $\rho$ describe the vectors of measurements being correlated. We plot the results as a correlation matrix in Fig. 1B, where we can see relatively high correlations between most of the variables analyzed. The high correlations we see are consistent with associations seen in a multitude of previous economic and sociological studies \cite{kahl1955comparison,lawson1960correlations,ludwig2001urban,moller2009changing}, although our framework has the added benefit of utilizing a single unified formalism to analyze all these associations. However, to get at underlying universal mechanisms behind observed socioeconomic data, we must go beyond solely demonstrating statistical associations between variables. The correlations seen in Fig. 1B may actually just be an artifact of a more fundamental process determining the decays in Fig. 1A, and we can make some headway in uncovering this process (or processes) using techniques inspired from urban scaling.\\

Traditional urban scaling posits that a wide variety of characteristics $Y$ in a city can be predicted solely by the city's population $P$ through relations of the form $Y\sim P^{\eta}$ for some exponent $\eta >0$, which in practice holds up for a large number of cities and variables of interest \cite{bettencourt2007growth}. The success of the urban scaling theory relies on a few key factors that are associated with a growing city population: denser organization of facilities and infrastructure, an accelerated pace of life, and increased interaction between agents and activities leading to specialization and innovation \cite{bettencourt2010unified}. In practice, the data $Y$ for some city-level characteristic (such as new patents or number of gas stations) is fit versus city population $P$ for many different cities, yielding an estimate for the exponent $\eta$ which we can interpret to gain an understanding of the fundamental processes contributing to the scaling behavior of $Y$. For instance, if $\eta >1$ this says that $Y$ grows superlinearly with $P$, which should be the case for quantities $Y$ that show increasing returns with population (e.g. indicators of innovation such as new patents). On the other hand, $\eta < 1$ indicates economies of scale, or characteristics $Y$ that decrease in unit cost as we increase the city's population (e.g. mobility-related infrastructure such as number of gas stations). Perhaps the most important takeaway from traditional urban scaling analysis is that when we can collapse the behavior of a wide range of seemingly different socioeconomic systems into a single framework with few parameters, these parameters can help us understand basic universal processes underlying these superficially distinct variables.\\

We can use a similar process to interpret the results of Fig. 1A, except rather than the absolute quantity of a socioeconomic indicator, we are analyzing correlations between distribution-valued quantities, and the fundamental covariate here is network distance $l$ instead of population $P$. Based on their homogeneous populations and degrees, the total population in tracts at path distance $l$ or less from a focal tract is very similar across tracts, and so $l$ reflects the total population included as we encircle a focal tract at greater and greater radii. As space is a factor for socioeconomic processes mainly in that it provides a medium for interaction among people \cite{redding2017quantitative}, this distance $l$ may be a more fundamental quantity than standard spatial distance in how it determines socioeconomic activity, and so we may be able to explain the spatial variation in a wide variety of socioeconomic variables using simple functions of $l$ such as Eq. \ref{pl_form}. An alternative quantity to $l$ could be derived from literally transforming space based on population to homogenize the population density, a concept which has inspired numerous interesting and informative mapping methods \cite{gastner2004diffusion}. However, we are ultimately constrained by the basic spatial units designated for data aggregation (e.g. census tracts), and so here we treat these regions, hence $l$, as fundamental.  \\

In the present case, we can see that the exponents $\beta$ determining the network correlation cutoff scale are very similar for education, housing, income, and race, indicating that correlations in these regional distributions are non-negligible over a universal distance scale of $\sim 30$ hops. However, we see higher variation in the scaling exponents $\alpha$, with race and housing decaying at a slower rate across the network than education and income. This suggests that perhaps the mechanisms that drive spatial differences in racial composition and local real estate values operate over larger distances than the mechanisms behind income or educational variability, at least in the US. \\

The association between the spatial distributions of housing values and racial groups has been noted in numerous studies that address ``redlining" and other processes that result in lower property values in neighborhoods with high minority populations \cite{zenou2000racial}. The analyses in Fig. 1A may point to additional, more subtle mechanisms behind this inequality due to a significant difference in the scaling exponents for housing and race, as this observed discrepancy indicates that the scales over which housing and racial regional similarity decay are quite different. It is known that home values are also tied to local incomes, which in turn can result in high variability in housing prices due to the relative flexibility of wages and mobility of workers compared to supply-regulated housing \cite{van2010has}. Therefore, perhaps the interplay between the long-range correlated racial composition of the population and the comparatively short-range correlated income distributions play a role in determining the moderate decay exponent $\alpha$ we see in the data. However, more definite conclusions and practical intervention strategies require a more contextualized analysis in conjunction with domain expertise.

\begin{figure}
\label{fig:figure1}
    \centering
    \includegraphics[width = 1\textwidth]{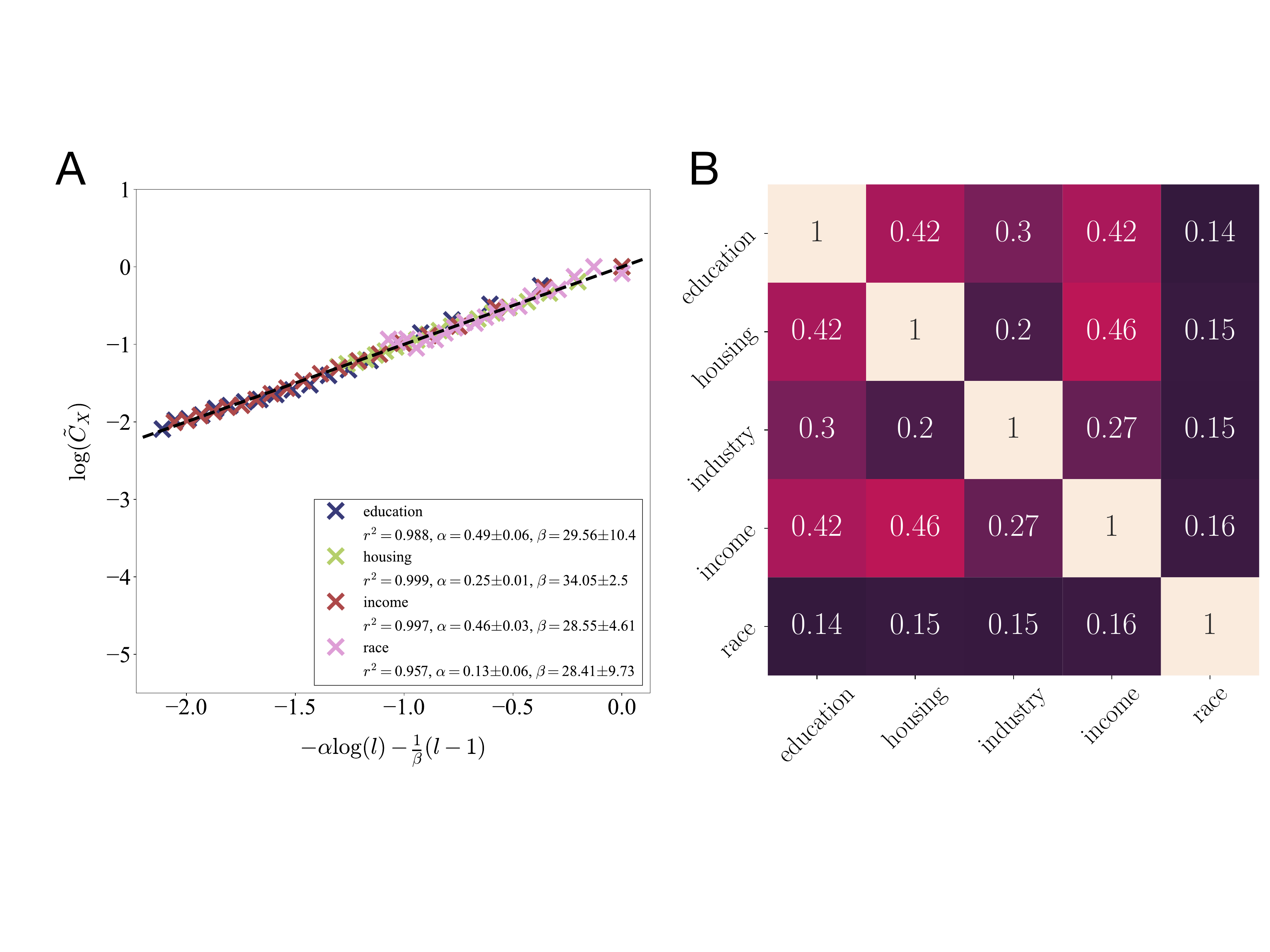}
    \caption{\textbf{Universal patterns in tract similarity across attributes.} \textbf{(A)} Fit results for the two-point correlation functions $C_X(l)$ for attributes $X$ in Table I, with $95\%$ confidence intervals around the scaling and cutoff exponents $\alpha$ and $\beta$. The line $y=x$ is plotted for reference, as a perfect scaling collapse maps all points onto this line. Eq. \ref{fit} is deemed a poor fit for $\tilde C_{industry}$ after a residual analysis, and so this result is omitted. \textbf{(B)} Spearman correlation matrix with respect to losses $L^{(ij)}_X$ across all edges $(i,j)$, for all pairs of socioeconomic attributes utilized in our study. All correlations are highly statistically significant at the $1\%$ significance level, with standard errors of $\sim 0.001$.}
\end{figure}

\subsection{County-level Heterogeneity}

To examine the regional diversity of a given socioeconomic variable, we employ Eq. \ref{L_eq}, this time to all the tracts comprising each county within our dataset. More specifically, for each county we examine, we compute $L^{(t_1,t_2,...)}_X$ with $t_i$ the census tracts within the county subdivision and $X$ the variable of interest. For notational convenience, we will use the notation $L^{(county)}_X$ from now on for this quantity. In order to compare counties with varying numbers of constituent tracts on the same scale, we normalize for potential biases from the number of included tracts by using a bootstrapping procedure to compute z-scores for each county-level value $L^{(county)}_X$. To do this, for all county sizes (number of constituent tracts) $k$  we compute the vectors $\mu_k$ and $\sigma_k$, which are the sample mean and standard deviation of $L^{(t_1,...,t_k)}_X$ over 100 random samples of $k$ tracts ${t_1,...,t_k}$. Then, we calculate a standardized version of Eq. \ref{L_eq}, $\tilde L$, for each county using
\begin{align}
\label{eq:L_county}
\tilde L^{(county)}_{X} = \frac{L^{(county)}_{X}-\mu_{\abs{county}}}{\sigma_{\abs{county}}},
\end{align}
where $\abs{county}$ is the number of tracts within the county. We will refer to Eq. \ref{eq:L_county} as a ``disparity" measure, as higher values of $\tilde L^{(county)}_{X}$ indicate higher dissimilarity in a county's tract-level distributions of $q^{(i)}_X$ relative to what is expected in a randomized null model where the county's tracts are chosen at random. In practice, we will see that $\tilde L$ tends to be negative for most counties, and in this case we should note that values of greater magnitude indicate higher \emph{similarity} in the county-aggregated tracts than expected by chance. \\

As a first step in understanding county-level disparities across the US, we plot the distribution of $\tilde L^{(county)}_X$ over all counties for each socioeconomic attribute $X$ in Fig. 2A as box-and-whisker plots. We can see that the distributions of all quantities tend strongly towards negative values, indicating that most counties have greater similarity in their tract-level distributions $q^{(i)}_X$ than expected in the null model. This is consistent with the spatial autocorrelation at short scales we see in socioeconomic variables in Fig. 1A, although these analyses in some sense provide a complimentary view point. Here, rather than assessing the scales over which distributions of socioeconomic characteristics remain similar as in Fig. 1, we are examining whether artificially drawn administrative boundaries are effective at capturing the homogeneity in these attributes. As counties have size scales much smaller than the area covered up to the typical correlation cutoff scale $l\sim 30$ from any reference tract, we expect that correlations between tract-level distributions will be relatively high within counties, and so in this sense these results should be unsurprising. \\

Looking at Fig. 2A, we do see something perhaps unexpected though: there are lots of counties that are only slightly more (and sometimes less) homogeneous in their tract-level distributional data than we'd expect by chance. In particular, most of the values of $\tilde L^{(county)}_{race}$ are in the interval $[-2,0]$, which means they are less than two standard deviations different in their disparity than expected with completely randomized tracts. This suggests that many counties in the US are relatively representative of the whole US in terms of racial composition, whereas there are relatively few counties with drastically different compositions. The same does not hold for housing though, for which around $75\%$ of the counties studied had more than two standard deviations differentiating their disparity values from the null model expectation. This result indicates that there are relatively few counties with distributions of housing values that are diverse enough to reflect typical housing prices nationally.   \\

To determine the association in the disparity values $\tilde L^{(county)}_X$ across counties, we plot the corresponding Spearman correlation matrix using the results from all counties studied. Similarly to Eq. \ref{rho}, we compute
\begin{align}
\label{spearman_counties}
\rho\left(\{\tilde L_X^{(c_1)}:c_1\in \text{counties}\},\{\tilde L_X^{(c_2)}:c_2\in \text{counties}\}\right).
\end{align}
The Spearman correlation matrix in Eq. \ref{spearman_counties} is shown in Fig. 2B, where we can see very high correlations between the within-county disparities, even higher than in the values of $L^{(ij)}_X$ shown in Fig. 1B. These correlations are similar in sign and relative magnitude (between attributes $X$) to those in Fig. 1B, but by aggregating tracts at the county level rather than just assessing correlations over edges, we are effectively reducing noise by smoothing out local fluctuations, and so we see a major increase in the values of $\rho$. In other words, some individual edges $(i,j)$ may have very different divergences $L^{(ij)}_X$ and $L^{(ij)}_{X'}$, but the effect of these outlier pairwise relationships is reduced when looking at distributions between tracts at the county-level. This noise reduction is only possible because, as discussed, the scale at which we are analyzing $\tilde L^{(county)}_X$ is smaller than the area associated with the correlation cutoff scales $\beta$ found in Fig. 1A.\\

Finally, as a case study to visualize the geographic manifestation of these county-level disparities, we plot a heatmap of $\tilde L^{(county)}_{housing}$ across all counties studied in Fig. 2C. Here we can immediately see an interesting pattern: the county-level disparity in housing prices, when compared to the same number of randomly selected regions, is actually much \emph{lower} along the coasts and metropolitan areas than it is elsewhere. Housing markets in coastal and metropolitan regions are typically seen as having high inequality due to the large variation in home and land values often seen in these areas \cite{dwyer2007expanding,davis2008price}. However, when assessed on a relative scale using distributions at the granularity of census tracts, we see a different story. In this case, we see that these coastal and metropolitan counties actually have quite similar distributions $q^{(i)}_{housing}$ across their constituent tracts $i$ relative to more inland and rural counties. The primary reason for this may be that the heterogeneity in housing prices in dense, urban counties is primarily manifested at scales below our measurement precision: tracts themselves have house price distributions with high variance, but tracts in a given county tend to have relatively similar distributions. This is consistent with the low rate of spatial decay in housing correlations seen in Fig. 1A, as most tracts are urban \cite{wang2013population} and if most of the fluctuations persist at scales smaller than census tracts, we will see a relatively smooth correlation trend at larger scales. Due to the coarse binning of housing values, however, there is also a potential confounding factor here in that many of the home prices in expensive metropolitan and coastal regions fall into the highest bin in the corresponding census data ($>\$1,000,000$), and so variability due to fluctuations above this price threshold are suppressed when using census data to assess inequalities. 

\begin{figure}
\label{fig:figure2}
    \centering
    \includegraphics[width = 1\textwidth]{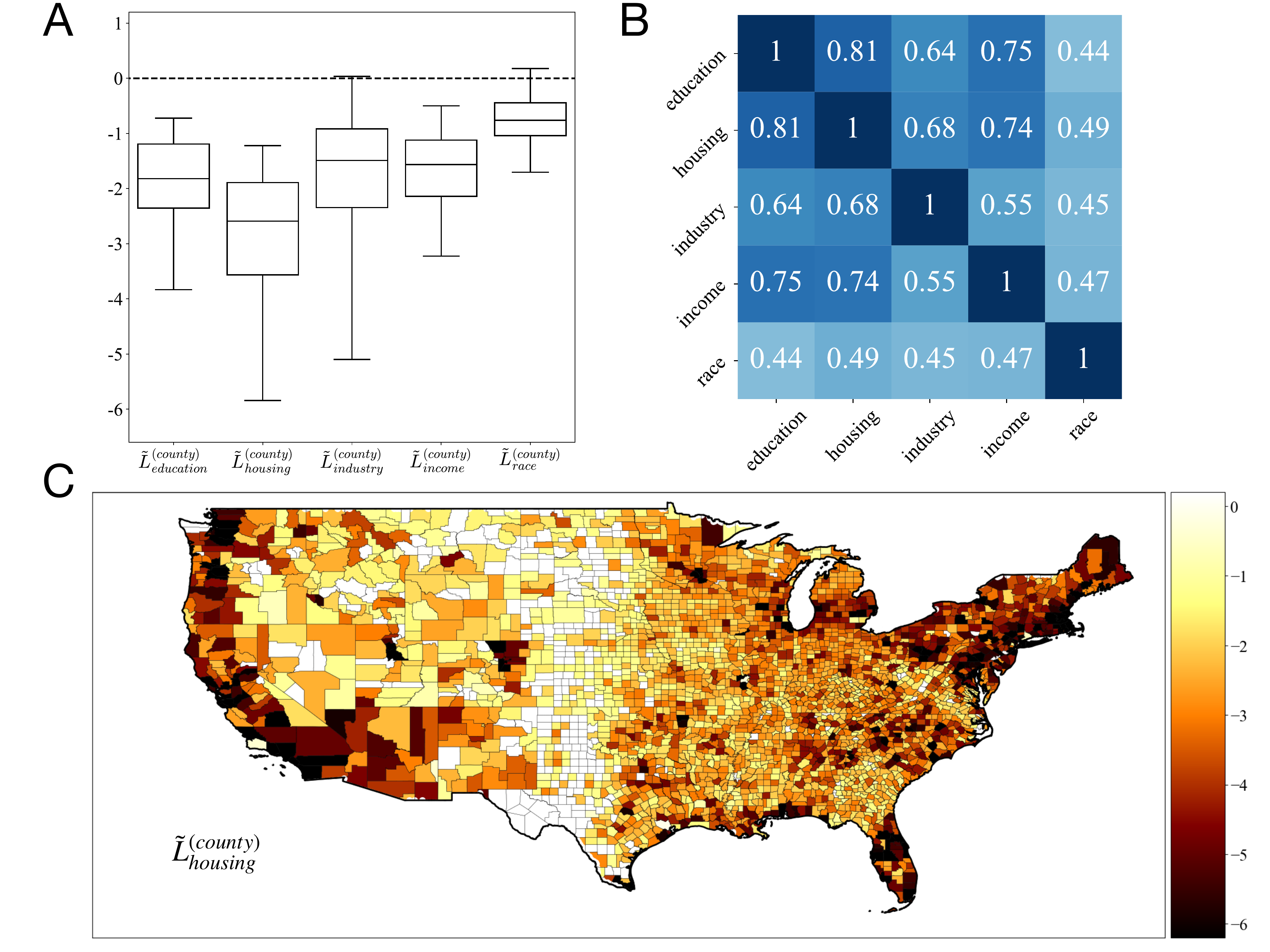}
    \caption{\textbf{County-level distributional disparity.} \textbf{(A)} Distribution of $\tilde L_X$ for all measures $X$, showing trends towards negative values indicating lower within-county disparity than expected by chance in a randomized null model. Whiskers extend to the $5$th and $95$th percentiles. \textbf{(B)} Spearman correlation between county-level disparity measures $\tilde L^{(county)}_X$ for all pairs of attributes, showing a high positive association between all pairs of variables. All correlations are highly statistically significant at the $1\%$ significance level, with standard errors of $\sim 0.01$. \textbf{(C)} Housing price disparities $\tilde L^{(county)}_{housing}$ for all counties in the continental United States. Values of $\tilde L$ indicate the degree to which counties' within-county distributional similarity in housing values differs from expectation (with more negative values associated with very high within-county similarity compared to expected).}
\end{figure}

\subsection{Regional Clustering}

As a final experiment using our measures, we detect communities at multiple size scales in the census tract subnetwork within the city of Chicago---a frequently used case study in socioeconomic diversity due to its rich history and abundance of available data \cite{doussard2009after,kassen2013promising}---with the goal of constructing clusters that are relatively homogeneous with respect to each socioeconomic attribute. Optimal aggregation of spatial regions according to various criteria has been a longstanding problem of interest, and numerous approaches have been proposed to tackle this using networks with edges weighted by an attribute representing regional similarity. This approach has the added benefit that since community detection algorithms look for connected clusters of nodes, the clusters detected naturally tend to be contiguous, and thus relevant for spatially localized policy. Attributes used in previous studies include phone calls between regions \cite{sobolevsky2014general}, commuting flows \cite{nelson2016economic}, taxi trips \cite{liu2015social}, and similarity between individual within-region features like our own method \cite{assunccao2006efficient}. \\

In order to group the tract network into clusters that exhibit homogeneity with respect to attribute $X$, we use $L^{(ij)}_X$ to construct edge weights $w_{ij}$ for the network prior to performing community detection. However, we cannot not use $L^{(ij)}_X$ for edge weights directly, as community detection algorithms typically associate higher edge weight with higher node similarity, and $L^{(ij)}_X$ is structured so that \emph{lower} values indicate greater similarity across an edge $(i,j)$. We thus employ a common transformation from the machine learning literature \cite{kondor2002diffusion}, which is to use an exponential kernel to map the values $L^{(ij)}_X$ to their associated edge weights $w_{ij}$ in the network. The weight transformation can be written as
\begin{align}
w_{ij} = e^{-\omega L^{(ij)}_X},    
\end{align}
where $\omega > 0$ is a tunable parameter that determines how much differentiation in the weights we will have across edges in the network. A value of $\omega \approx 0$ results in almost no differentiation between edge weights ($w_{ij}\approx 1$ for all edges), whereas $\omega >> 1$ results in an exaggerated difference in edge weights between edges with lower $L^{(ij)}_X$ and edges with higher $L^{(ij)}_X$. Any kernel mapping the unit interval to decreasing non-negative reals would suffice to construct the weights $w_{ij}$, but we opt for the exponential function here because it is particularly simple and only has one tunable parameter. For the experiments shown, we find a middle ground between the two extremes presented for $\omega$, for each attribute-based clustering choosing a value of $\omega$ that results in a relatively uniform distribution of edge weights across $[0,1]$. More specifically, for each attribute $X$ we numerically approximate the $\omega$ that maximizes the entropy of the associated distribution of edge weights $e^{-\omega L^{(ij)}_X}$. A more principled method for choosing $\omega$ based on the application of interest is a subject is left to future work, but here we use this simple statistical procedure to avoid falling into one of the two cases presented, where there is either no differentiation in the edge weights or only a handful of edges matter. \\

In order to detect communities in the Chicago subnetwork, we aim to find the configuration of node communities $\vec{c}=\{c_i\}$ in the subnetwork such that the weighted modularity $Q_\gamma(\vec{c})$ is approximately optimized. The modularity $Q_\gamma(\vec{c})$ that we use here is defined by
\begin{align}
\label{mod}
Q_\gamma(\vec{c})=\frac{1}{W}\sum_{ij}\left[\gamma w_{ij}- \frac{s_is_j}{W}\right]\delta_{c_i,c_j},    
\end{align}
where $W$ is the sum of edge weights in the network, $s_i = \sum_{k}w_{ik}$ is the sum of weights of edges attached to node $i$, and $\gamma$ is a \emph{resolution parameter} \cite{reichardt2006statistical}. When $\gamma = 1$, Eq. \ref{mod} reduces to the traditional notion of weighted modularity, but varying $\gamma \neq 1$ allows us to choose the importance given to $w_{ij}$ relative to $\frac{s_is_j}{W}$ (which is the approximate expected weight of $w_{ij}$ through random rewiring). In particular, the larger we make $\gamma$, the more importance is given to the observed edge weights relative to the expected weights, and the community configurations $\vec{c}$ that maximize Eq. \ref{mod} will be larger. Thus, by varying $\omega$ we can tune how much influence differences in $L^{(ij)}_X$ across edges have, and by varying $\gamma$ we can determine the characteristic cluster size. We use the Louvain Algorithm \cite{blondel2008fast}, a greedy optimization method, to find the configuration $\vec{c}$ that approximately maximizes Eq. \ref{mod}. There are numerous viable alternative methods but here we opt for the Louvain algorithm as it is fast and straightforward to implement. It is also important to note that we can perform regional aggregation with $L^{(ij)}$ in a manner where clusters are not likely to be contiguous, for instance by constructing a matrix from all pairwise values of $L^{(ij)}$ and performing one of various matrix-clustering techniques \cite{jain1999data}. However, here we are interested in constructing contiguous clusters of tracts in order to coarse grain the city into zones relevant for spatially targeted policy intervention, and so we use community detection to encourage contiguity of the clusters. \\

In Fig. 3 we show the results of our community detection analysis for the Chicago census tract subnetwork. In Fig. 3A-3C, we show the clusters obtained for edge weights constructed using $L^{(ij)}_{income}$, at various resolutions $\gamma$. We can observe that increasing $\gamma$ allows us to get a coarser view of the socioeconomic clusters present in the city, and can allow for delineation of super-regions at a desired scale. We also show the officially designated neighborhood boundaries (thick black lines) for Chicago (\url{https://data.cityofchicago.org/}) in order to visualize the consistencies and inconsistencies between our clusters and these officially delineated regions. We can see that in the intermediate regime $\gamma\sim 0.1$, our clusters are consistent with some neighborhood boundaries, but deviate significantly from others. This suggests that the officially designated regions are somewhat consistent with homogeneous socioeconomic clusters, but there is room for improvement to these boundaries if the goal is to delineate socioeconomically homogeneous zones within the city (at least regarding income). Of course, there are numerous other factors, both socioeconomic and geographic, that would need to be accounted for in addition to the factors we analyze in order to draw effective policy-relevant boundaries in practice. \\

We also compute the Adjusted Mutual Information (AMI) between clusters obtained using different attributes $X$ as well as the official neighborhood clusters, in order to assess the consistency in the groups we obtain when considering these different factors. The Mutual Information $MI(\vec{c}_1,\vec{c}_2)$ is the amount of shared information (in an information theoretic sense) between the clusterings $\vec{c}_1$ and $\vec{c}_2$, or more intuitively, the statistical uncertainty in each independent clustering minus the statistical uncertainty when combined. More specifically, we have that
\begin{align}
MI(\vec{c}_1,\vec{c}_2) = H[\vec{q}_1]+H[\vec{q}_2]-H[\vec{q}_{12}] = \sum_{s,t}\vec{q}_{12}(s,t)\log\left[\frac{\vec{q}_{12}(s,t)}{\vec{q}_1(s)\vec{q}_2(t)}\right]    
\end{align}
where $\vec{q}_1(s)$ is the fraction of nodes put into cluster $s$ under configuration $\vec{c}_1$ (and similarly for $\vec{q}_2$), and $\vec{q}_{12}(s,t)$ is the fraction of nodes put into group $s$ under configuration $\vec{c}_1$ and $t$ under configuration $\vec{c}_2$. One drawback to using MI, however, is that it gives systematically higher values as we increase the number of clusters, even for completely random cluster configurations \cite{vinh2010information}. One proposed correction (of many) to this is to use the AMI, given by
\begin{align}
AMI(\vec{c}_1,\vec{c}_2) = \frac{MI(\vec{c}_1,\vec{c}_2) - \left\langle MI(\vec{c}_1,\vec{c}_2)\right\rangle}{\text{Max}(H[\vec{q}_1],H[\vec{q}_2])-\left\langle MI(\vec{c}_1,\vec{c}_2)\right\rangle},    
\end{align}
where $\left\langle MI(\vec{c}_1,\vec{c}_2)\right\rangle$ is the expectation value of MI in the null model where the number of items in each cluster is fixed and groups are generated randomly through permutations of labels. The AMI is equal to $0$ if the clusters $\vec{c}_1$ and $\vec{c}_2$ share the amount of information we expect from random chance purely based on their cluster sizes, and $1$ if the clusters are identical.\\

In Fig. 3D, we plot the average AMI over all pairs of partitions using the five socioeconomic attributes, as a function of the resolution parameter $\gamma$. We can see that there is a clear peak value of $\gamma$ at which the five attributes share highly overlapping clusters. In practice, this could be used as a heuristic to tune $\gamma$ for selecting the size scale of the clusters, if the goal is to select clusters that are highly homogeneous with respect to multiple socioeconomic attributes. It is interesting to note the clear scale sensitivity in this analysis: at certain scales, we can divide the city into zones that are relatively socioeconomically homogeneous in all variables studied, but at other scales, the city decomposes into regions with less overlap. \\  

Fig. 3E shows the AMI matrix for the clusters obtained at $\gamma\approx 0.1$, the peak in Fig. 3D. We can see from this plot that all socioeconomic attributes are spatially clustered in quite similar patterns at this scale, and that all have high correlation with the official neighborhood boundaries as well. This is perhaps an endorsement for the neighborhood boundaries, as these results suggest that the scale at which the neighborhoods are drawn corresponds to the scale at which the socioeconomic clusters in the city are most similar. Taken together, the results from Fig. 3D and 3E may point to a new method for subdividing a city into different neighborhoods, which can be constructed easily based on any socioeconomic attribute and at any size scale. \\

\begin{figure}
\label{fig:figure3}
    \centering
    \includegraphics[width = 1\textwidth]{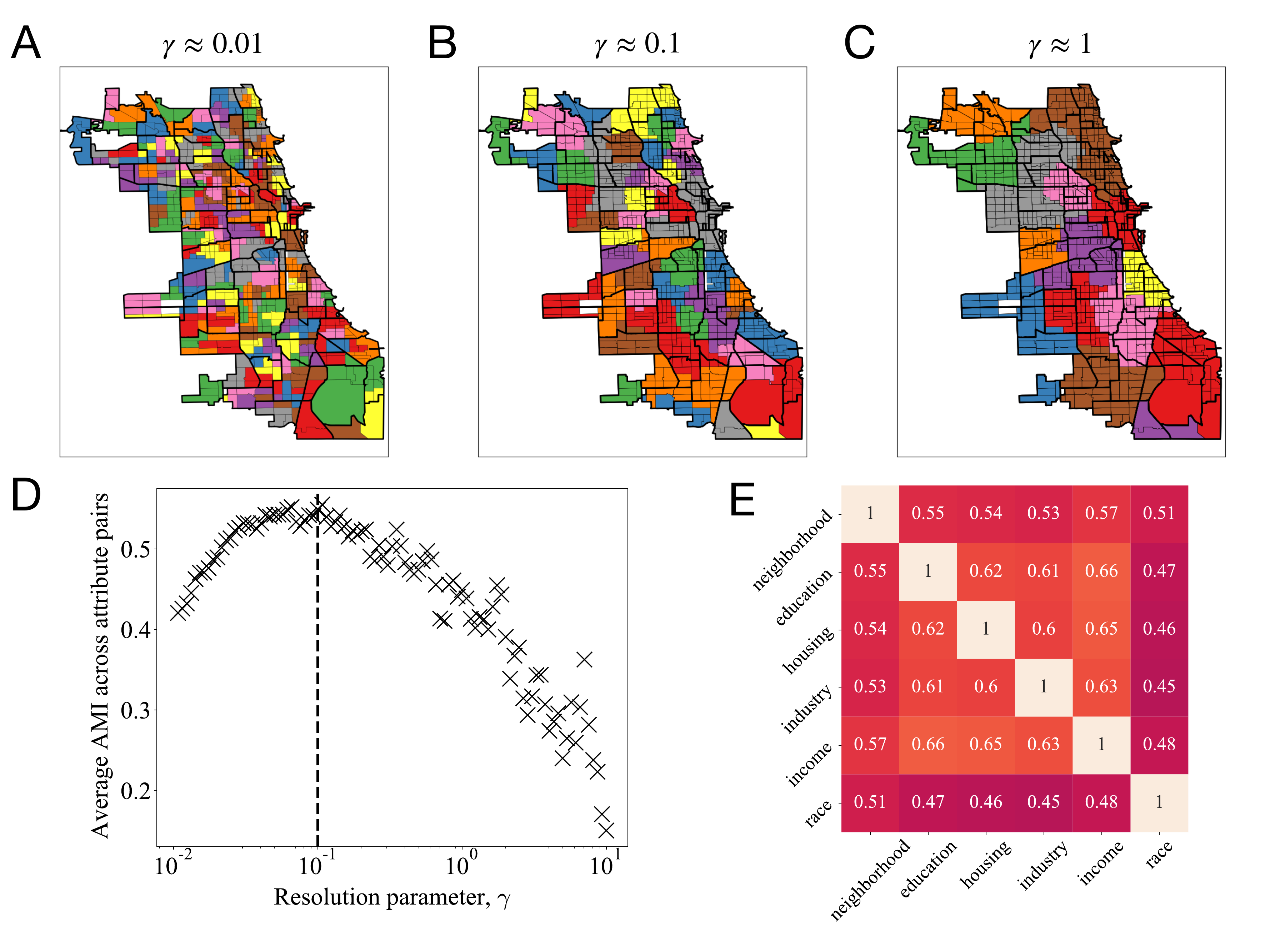}
    \caption{\textbf{Attribute-based regional clustering at multiple scales.} \textbf{(A)-(C)} Clusters obtained through weighted community detection for census tracts (thin black lines) in Chicago with respect to income for various resolution parameters $\gamma$, displaying varying characteristic size and association with neighborhood boundaries (thick black lines). \textbf{(D)} Adjusted Mutual Information (AMI) between the clusters obtained by our community detection algorithm, as a function of $\gamma$ and averaged over all pairwise combinations of the five studied socioeconomic variables. We see a clear peak value of $\gamma \approx 0.1$ at which the clusters obtained through the five methods are highly correlated (dashed vertical line). \textbf{(E)} AMI matrix between clusters computed at this peak $\gamma$ value, including the clusters obtained by grouping tracts by the neighborhood they most overlap with, indicating a high correlation between all of these partitions compared to what one expects by random chance based on their cluster sizes.}
\end{figure}

\section{Conclusion}

In this study, we propose a new measure for analyzing socioeconomic data across spatial regions using concepts from network theory and information theory, which accommodates all forms of distributional data, has a natural extension to the comparison of more than two regions, and allows for policy-relevant analysis by considering officially delineated regions as fundamental spatial units. By analyzing spatial data from a topological lens, we can approach regional analysis issues from a relational perspective that avoids the longstanding issue of identifying appropriate spatial scales. We apply our framework in a series of experiments on the adjacency network of US census tracts to demonstrate the new insights we can gain with our methodology. We first find a universal decay pattern in various socioeconomic correlations as a function of path distance, as well as high statistical association between distributional similarities in adjacent tracts. We then aggregate tract-level distributions at the county level, finding again that distributional disparity measures are highly correlated, and also that there are relatively low levels of within-county inequality compared to what one would expect by aggregation of random tracts. Finally, we propose a clustering algorithm for regional aggregation into homogeneous socioeconomic clusters, finding that in practice the clusters obtained by our methodology have high overlap with accepted neighborhood delineations, as well as with each other across attributes. These applications illustrate the versatility of our methods, as well as the universality present in socioeconomic data when analyzed with a unified framework. \\

There are numerous improvements that can be made to our methodology in future work that increase its effectiveness in practical applications. Firstly, important limitations arise from the quality and resolution of census data, which we do not attempt to address as they are outside the scope of this work. In particular, the coarse binning of interval distributional datasets (here, income and housing) can result in poor estimation of entropy and other uncertainty measures, as long tails are not accounted for, which may account for a large portion of the variability in the distributions \cite{ferson2007experimental}. One improvement to our methodology to obtain more accurate results would thus be to estimate these full distributions based on the predefined bins and other summary statistics such as mean, median, and Gini coefficient \cite{von2016robust,von2017better}, then apply our measures using approximations of  differential entropy. Additionally, some census data have large margins of error due to various statistical sampling issues \cite{national2007using,spielman2014patterns}, and so correcting for this noise in our analyses would also improve the efficacy of our techniques. However, we leave these and further improvements to future work.

\begin{acknowledgments}
The author thanks Shihui Feng and Mark Newman for helpful discussions. A.K. was supported by the Department of Defense through the National Defense Science and Engineering Graduate Fellowship (NDSEG) program. 
\end{acknowledgments}


\onecolumngrid
\begin{center}
\begin{table}[h]
\begin{tabular}{|c|c|c|}
\hline
\textbf{variable }$X$ & \textbf{support $\{x\}$ of }$q_X$ &
\textbf{ACS codes}\\ 
\hline

race & \begin{tabular}{c}
White \\
Black or African American\\
American Indian and Alaska Native\\
Asian\\
Native Hawaiian and Other Pacific Islander\\
Other
\end{tabular} & DP05, 0059PE - 0064PE \\ \hline 

income & \begin{tabular}{c}
Less than 10,000 \\
10,000 - 15,000\\
15,000 - 25,000\\
25,000 - 35,000\\
35,000 - 50,000\\
50,000 - 75,000\\
75,000 - 100,000\\
100,000 - 150,000\\
150,000 - 200,000\\
Greater than 200,000
\end{tabular} & DP03, 0052PE - 0061PE \\ \hline

industry & \begin{tabular}{c}
Agriculture, forestry, fishing and hunting, and mining \\ 
Construction\\
Manufacturing\\
Wholesale trade\\
Retail trade\\
Transportation and warehousing, and utilities\\
Information\\
Finance and insurance, and real estate and rental and leasing\\
Professional, scientific, management, and administrative services\\
Educational services, and health care and social assistance\\
Arts, entertainment, recreation, accommodation, and food services\\
Other services, except public administration\\
Public administration
\end{tabular} & DP03, 0033PE - 0045PE \\ \hline

housing (value) & \begin{tabular}{c}
Less than 50,000 \\
50,000 - 100,000\\
100,000 - 150,000\\
150,000 - 200,000\\
200,000 - 300,000\\
300,000 - 500,000\\
500,000 - 1,000,000\\
Greater than 1,000,000
\end{tabular} & DP04, 0081PE - 0088PE \\ \hline

education & \begin{tabular}{c}
Less than 9th grade \\
9th to 12th grade, no diploma\\
High school graduate (includes equivalency)\\
Some college, no degree\\
Associate's degree\\
Bachelor's degree\\
Graduate or professional degree 
\end{tabular} & DP02, 0059PE - 0065PE \\ 
\hline
\end{tabular}

\caption{Information on ACS distributional variables. For each variable $X$, we show its support as well as the associated ACS variable codes from \url{https://api.census.gov/data/2018/acs/acs5/profile/variables.html}.} 
\label{ACSvars}
\end{table}

\end{center}
\twocolumngrid

\end{document}